\documentclass[
aps,
prb,
preprint,
superscriptaddress,
longbibliography,
amsmath,
amssymb,
floatfix.
xcolor=table]{revtex4-1}
\usepackage{graphicx}
\usepackage{dcolumn}
\usepackage{bm}

\usepackage[utf8]{inputenc}
\usepackage[T1]{fontenc}
\usepackage{mathptmx}
\usepackage{subcaption}
\usepackage[mathlines]{lineno}
\usepackage{siunitx}
\usepackage[table,xcdraw]{xcolor}
\usepackage{booktabs}
\usepackage[makeroom]{cancel}

\begin{document}
	\title[Dependence of energy barrier reduction on collective excitations in square artificial spin ice]{Dependence of energy barrier reduction on collective excitations in square artificial spin ice: A comprehensive comparison of simulation techniques}

	\newcommand{\univie}{Christian Doppler Laboratory, “Advanced Magnetic Sensing and Materials”, University of Vienna, Waehringer Strasse 17, 1090 Vienna, Austria}
	\newcommand{\PSIMesosys}{Laboratory for Multiscale Materials Experiments, Paul Scherrer Institute, 5232 Villigen PSI,~Switzerland}
	\newcommand{\ETHMesosys}{Laboratory for Mesoscopic Systems, Department of Materials, ETH Zurich, 8093 Zurich,~Switzerland}
	\newcommand{\nanogune}{CIC nanoGUNE BRTA, 20018 Donostia-San Sebastián, Spain}
	\newcommand{\ikerbasque}{IKERBASQUE, Basque Foundation for Science, 48013 Bilbao, Spain}
	\newcommand{\stockholm}{Department of Physics, Stockholm University, 106 91 Stockholm, Sweden}

	\author{Sabri~Koraltan}
	\email{sabri.koraltan@univie.ac.at}
	\affiliation{\univie}
	
	\author{Matteo~Pancaldi}
	\affiliation{\stockholm}
	
	\author{Na{\"e}mi~Leo}
	\affiliation{\nanogune}

	\author{Claas~Abert}
	\author{Christoph~Vogler}
	\affiliation{\univie}
	
	\author{Kevin~Hofhuis}
	\affiliation{\ETHMesosys}
	\affiliation{\PSIMesosys}
	
	\author{Florian~Slanovc}
	\author{Florian~Bruckner}
	\author{Paul~Heistracher}
	\affiliation{\univie}

	\author{Matteo Menniti}
	\affiliation{\nanogune}
	
	\author{Paolo~Vavassori}
	\affiliation{\nanogune}
	\affiliation{\ikerbasque}
	
	\author{Dieter~Suess}
	\affiliation{\univie}
	\date{\today}
	\preprint{AIP/123-QED}
	\begin{abstract}
		We perform micromagnetic simulations to study the switching barriers in square artificial spin ice systems consisting of elongated single domain magnetic islands arranged on a square lattice.
		By considering a double vertex composed of one central island and six nearest neighbor islands, we calculate the energy barriers between two types of double vertices by applying the string method.
		We investigate by means of micromagnetic simulations the consequences of the neighboring islands, the inhomogeneities in the magnetization of the islands and the reversal mechanisms on the energy barrier by comparing three different approaches with increasing complexity.
		The micromagnetic models, where the string method is applied, are compared to the currently common method, the mean barrier approximation. 
		Our investigations indicate that a proper micromagnetic modeling of the switching process leads to significantly lower energy barriers, by up to 35\% compared to the mean-barrier approximation, so decreasing the expected average life time up to seven orders of magnitude. Hereby, we investigate the influence of parallel switching channels and the conceptional approach of using a mean-barrier to calculate the corresponding rates.
	\end{abstract}
	\maketitle
	\section{\label{sec:level1}INTRODUCTION}
	Artificial spin ice (ASI) systems are lithographically patterned lattices of elongated single domain magnetic islands~\cite{Wang2006,nisoli_coll,Heyderman2013,Skjaervo2020}.
	In 2-dimensional ASI systems, the magnetic islands are arranged in vertices and build a frustrated lattice due to the competing magneto-static interactions among the islands~\cite{Wang2006, Nisoli2017, frustrations_ising,Kapaklis2014,moeller2006}.
	
\begin{figure}[]                                       
		\includegraphics[width=\columnwidth]{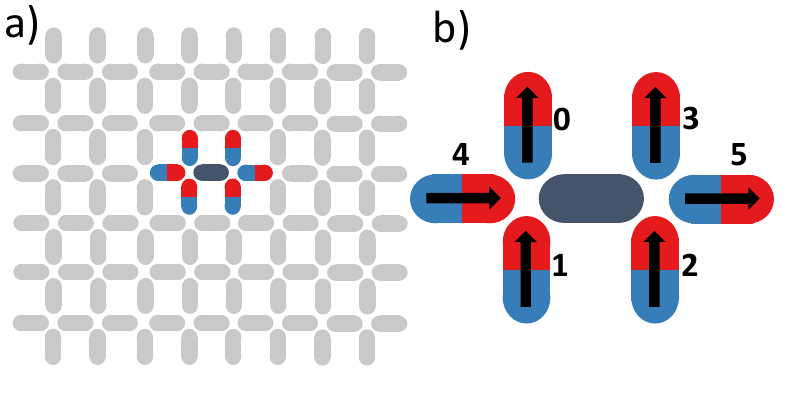}
		\caption{\textbf{Square Artificial Spin Ice lattice.} a)~Schematic illustration of a sASI lattice with an highlighted double vertex~(color). b)~Schematic illustration of the double vertex from a). With the central island~(dark), being the island of interest in the energy barrier calculations and the neighbouring islands enumerated with the island identification number $j$ to calculate the configuration number, as given later by Eq.~(\ref{eq:c_i}). }
		\label{fig:lat_dv}
	\end{figure}

\begin{figure}[]                                       
		\includegraphics[width=\columnwidth]{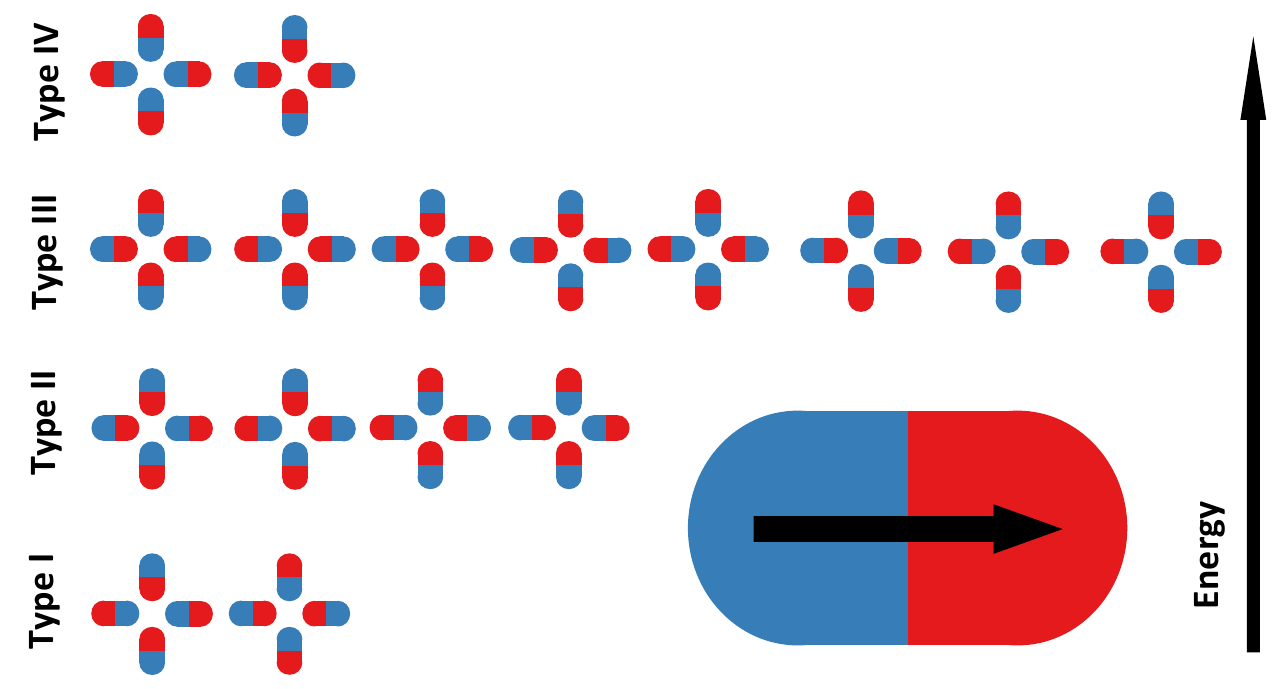}
		\caption{\textbf{Vertex configurations in a sASI.} Possible arrangements of the magnetizations of the islands in a square vertex, where the black arrow of the large island indicates the direction of the magnetization. All configurations that are assigned to the same vertex type have equal energies.}
		\label{fig:vertices}
	\end{figure}

	A possible ASI system, where four islands build a vertex and are arranged on a square lattice, is called square artificial spin ice~(sASI)~\cite{Wang2006,nisoli_coll}.
	Such a lattice is illustrated in Fig.~\ref{fig:lat_dv}a.
	In the absence of external magnetic fields, the magnetic islands are magnetized along their long axis due to the shape anisotropy, which limits the possible configurations in a vertex to $2^4=\num{16}$ macroscopic configurations.
	Fig.~\ref{fig:vertices} shows the four types of vertex configurations in a sASI, where the energy of each vertex increases with its assigned number.
	Type IV vertices contain magnetic islands with the magnetization pointing to the center or away from it.
	This type has the highest energy level and hence is the most excited configuration.
	Type~II and Type~I vertices obey the ice rule~\cite{nisoli_coll}, where two magnetizations are pointing to the center and two away from it.
	The ground state is represented by Type~I, since in a sASI lattice with alternating Type~I vertices the stray fields between four magnetic islands arranged in a square are building closed loops~\cite{Remhof2008}, minimizing the total energy of the system.

	Both experiments~\cite{Kapaklis_2012} and simulations~\cite{Budrikis2012, Greaves2012} have been performed to analyze the thermal annealing and domain dynamics in these structures.
	Furthermore, it has been experimentally observed that the ground state can be achieved over thermal relaxation, after the lattice has been subject to high temperatures~\cite{Zhang2013,Porro2013,farhan_prl,Kapaklis2014}.
	With the increasing temperature, the magnetic moments of the islands start to fluctuate. 
	For high enough temperatures, the islands may switch due to thermal activation, enabling the system to evolve towards the energetically favored ground state with Type I vertices.
	The temporal evolution of the thermally activated relaxation can be described by kinetic Monte Carlo simulations~\cite{charap97,farhan_prl,Thonig2014} that require the energy barriers of the various switching processes as an input. Thus the ability to accurately determine the energy barriers to be overcome in order to switch the magnetization of the magnetic islands are of key importance.
	By using the mean-barrier approximation to calculate the switching barriers, the experimentally observed relaxation mechanisms, and thus the dynamics and the ordering of the system, can only be artificially reproduced~\cite{Farhan2013,farhan_prl,farhanderlet2017,arava2019,Thonig2014}.
	Namely, in order to reproduce the relaxation timescale observed experimentally, the energy barriers need to be artificially reduced.
	Such a reduction is usually ascribed to extrinsic factors like fabrication defects, reduction of the Curie Temperature $T_C$~\cite{zhang2019} or saturation magnetization $M_s$~\cite{Kapaklis2014}.
	
	In this work, we study the dependence of the energy barriers on collective excitations in a square lattice by performing micromagnetic simulations.
	We apply the simplified and improved string method to investigate if the commonly reported barrier reduction is ascribable, and to what extent, to intrinsic fundamental physics related to the thermally induced reversal rather than a mere effect of extrinsic factors by comparing three different models.
	We find that these reductions can be ascribed to the preferential switching direction of the magnetizations in a magnetostatic environment, as well as non-uniform contributions to the moment reversal.
	\section{\label{sec:level2}MICROMAGNETICS}
	\subsection{\label{sec:level2_A}SIMPLIFIED AND IMPROVED STRING METHOD}
\begin{figure}[]   
		\includegraphics[width=\columnwidth]{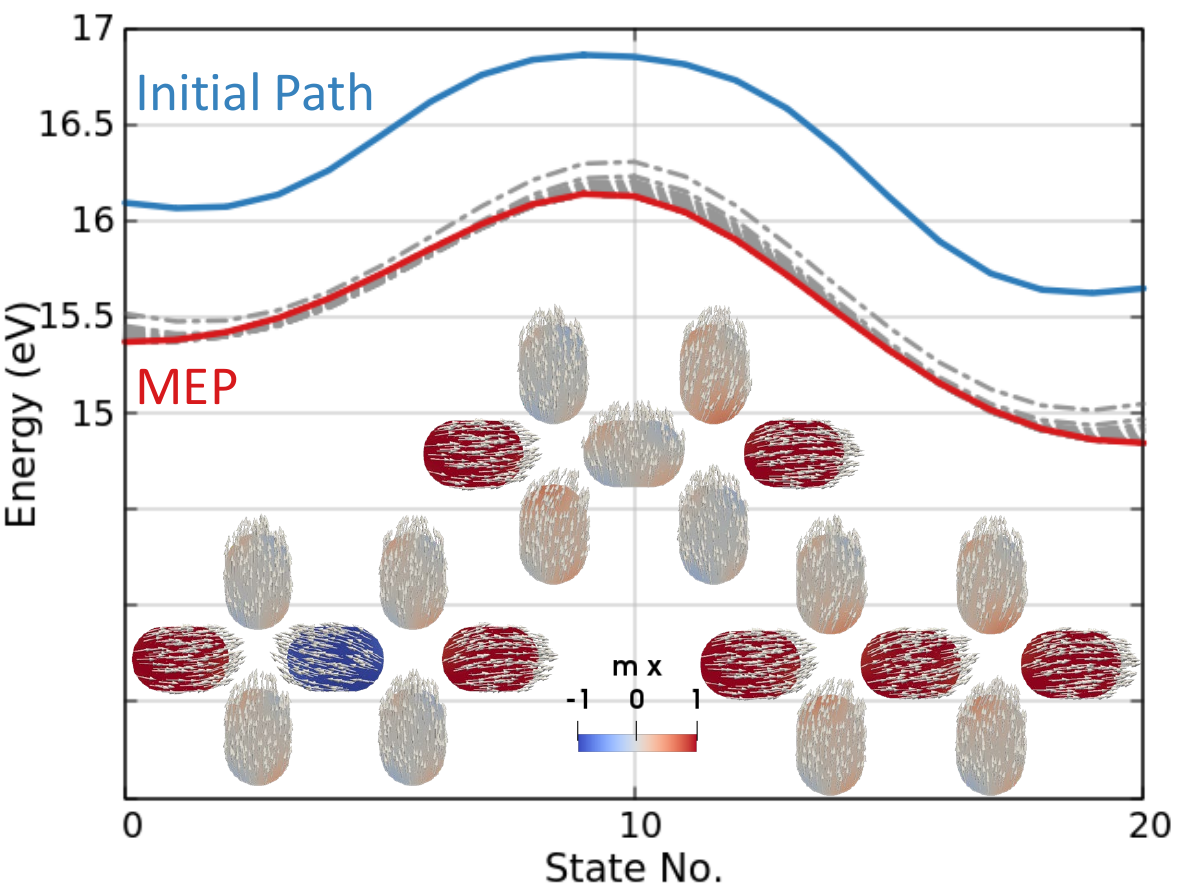}
		\caption{\textbf{Evolution of the energy paths during the SISM.} For the configuration $c=63$ illustrated in Fig.~\ref{fig:lat_dv}b, the initial path~(blue), represented by a coherent rotation, evolves towards the MEP~(red) via the intermediate energy paths~(gray). Inset figures represent the magnetization states for State~No.=0~(left), State~No.=10~(center) and State~No.=20~(right) of the MEP.}
		\label{fig:ep}
	\end{figure}

	In order to analyze the dependence of collective excitations on the energy barriers, we use the simplified and improved string method~(SISM)\cite{string} to calculate the energy barrier between two magnetic states.
	This method consists of three steps.
	In the first step, we create an initial path for the switching event.
	For this purpose, we choose a coherent rotation with the uniform magnetization $\boldsymbol{m}_{\mathrm{ini}}$ as our initial state. 
	The final state of the initial path is represented by $\boldsymbol{m}_{\mathrm{fin}}$.
	This path is discretized by 21 distinct magnetization configurations, which sample the continuous transition path in an equidistant fashion with respect to an appropriate norm \cite{string}.
	In a second step, each of these 21 magnetization configurations is evolved a certain amount towards its nearest energetic minimum, and in a final step, the 21 magnetization states are rearranged along the path in order to restore the equidistant discretization of the path.
	The last two steps are repeated until we obtain the minimum energy path~(MEP).
	This final path represents the lowest energy path and equally, the most favorable way to switch between  initial and final magnetization states with respect to the given initial path.
	Fig.~\ref{fig:ep} is an exemplary illustration for the evolution of the energy paths to obtain the MEP.
	In agreement with transition-state theory\cite{tst,Coffey2012} the energy barrier to switch the magnetization from $\boldsymbol{m}^{\mathrm{ini}}$ to $\boldsymbol{m}^{\mathrm{fin}}$ is obtained by
	\begin{equation}
	\centering
	\label{eq:de_tst}
	{\Delta E} =  {E}^{\mathrm{saddle}}-{E}^{\mathrm{ini}},
	\end{equation}
	where ${E}^{\mathrm{saddle}}$ is the energy corresponding to the saddle point of the MEP and $E^\mathrm{ini}$ is the energy of the initial state of the MEP.
	\subsection{\label{sec:level2_B}ENERGETICS}
	According to the string method, we minimize the total energy of a chosen magnetic region $\Omega_m$.
	For our purposes, we consider the total energy by means of micromagnetics~\cite{Abert2019} as
	\begin{equation}
	\centering
	\label{eq:etot_single}
	{E}^{\mathrm{tot}} =  {E}^{\mathrm{dem}}+{E}^{\mathrm{ex}},
	\end{equation}
	where ${E}^{\mathrm{dem}}$ denotes the demagnetization energy and ${E}^{\mathrm{ex}}$ represents the ferromagnetic exchange energy. We do not consider any externally applied magnetic fields.
	
	The demagnetization energy ${E}^{\mathrm{dem}}$ is described as
	\begin{equation}
	\centering
	\label{eq:e_dem}
	{E}^{\mathrm{dem}} = -\dfrac{\mu_0M_s}{2}\int_{\scriptsize{\Omega_{m}}}{{\boldsymbol{m}\cdot \boldsymbol{H}^{\mathrm{dem}}}\mathrm{d}\boldsymbol{x}},
	\end{equation}
	where $\mu_0$ is the vacuum permeability, $\Omega_m$ defines the magnetic region and $\boldsymbol{m} $ represents the normalized magnetization vector.
	
	The demagnetizing field $\boldsymbol{H}^{\mathrm{dem}}$ is given by
	\begin{equation}
	\centering
	\label{eq:h_dem}
	\boldsymbol{H}^{\mathrm{dem}}(\boldsymbol{x}) = -\frac{M_{\mathrm{s}}}{4\pi}\int_{\Omega_\mathrm{m}}{\nabla \nabla'\dfrac{1}{|\boldsymbol{x-x'}|}\boldsymbol{m}(\boldsymbol{x}')\mathrm{d}\boldsymbol{x}'}.
	\end{equation}
	Moreover, the ferromagnetic exchange energy favoring a parallel alignment of the spins is defined as
	\begin{equation}
	\centering
	\label{eq:e_ex}
	{E}^{\mathrm{ex}} = \int_{\scriptsize{\Omega_{m}}}{A^{ex}(\nabla{\boldsymbol{m}})^2\mathrm{d}\boldsymbol{x}},
	\end{equation}
	where $A^\mathrm{ex}$ is the exchange stiffness constant.
	\section{\label{sec:level3}MODELING}
	In this paper, we study the energy barriers between different vertex types of sASI.
	In particular, we show the influence of the nearest neighbor~(NN) islands on the energy barrier by computing the MEPs with different approaches.
	The considered models are summarized in Table~\ref{tab:models}.
	\begin{table*}[t]
		\caption{\textbf{Summary of the considered models.} The environment describes the magnetization of the NN islands and the central moment the magnetization of the central island during the switching event.}
		\begin{tabular}{cccc} 
			\toprule
			\toprule
			MODEL & MODEL~1 & MODEL~2 & MODEL~3\\  
			\midrule
			Description & Mean-barrier & Uniform environment & Full micromagnetic model\\
			Environment & Uniform & Uniform & Dynamically relaxed \\
			Central moment & Uniform & Nonuniform & Nonuniform\\
			Reversal mechanism & Coherent rotation & Minimum energy path & Minimum energy path\\            
			\bottomrule
			\bottomrule
		\end{tabular}
		\label{tab:models}
	\end{table*}
	To obtain a direct comparison between the models, we use \texttt{magnum.fe}\cite{magnumfe}, a finite element method based micromagnetic simulation code, to calculate the minimum energy path by applying the string method.
	We generate all finite element meshes using \texttt{Gmsh}\cite{gmsh}.
	
	We consider magnetic islands similar to those used to perform kinetic Monte Carlo simulations, for which the energy barriers needed to be reduced with respect to those arising using the geometrical and physical properties of the nanoelements to reproduce the experimentally observed relaxation. Our islands have a length L = \SI{150}{nm}, a width W = \SI{100}{nm}, an edge-to-edge gap g = \SI{90}{nm} and a thickness t = \SI{3}{nm}.
	Furthermore we use material parameters similar to bulk permalloy at $T~=~\SI{300}{K}$, with saturation magnetization $M_\mathrm{s}~=~\SI{790}{kA/m}$, exchange stiffness constant $A^{\mathrm{ex}}~=~\SI{13}{pJ/m}$ and vanishing uni-axial anisotropy constant $K~=~\num{0}$.
	
	In the following, we study the energy barrier for the switching event of the magnetization of the central island in a double vertex. Fig.~\ref{fig:lat_dv}b illustrates such a double vertex with the central island as the island of interest.
	In order to effectively label the 64 possible magnetization configurations of the NN islands, we introduce the integer parameter $c$ defined by
	\begin{equation}
	\centering
	\label{eq:c_i}
	c=\sum_{j=0}^{5}b_{j}2^{j}.
	\end{equation}
	
	The factor $b_j$ is obtained by
	\begin{equation}
	\centering
	\label{eq:bj}
	b_{j} = \begin{cases}
	0, & \text{if } m_k=-1\\
	1, & \text{if } m_k=+1
	\end{cases},
	\end{equation}  
	where $m_k$ is positive, if the magnetic moment points to the right~(up) for horizontal~(vertical) islands, as shown in Fig.~\ref{fig:lat_dv}b. The index $j$ denotes the NN island identification number as illustrated in Fig.~\ref{fig:lat_dv}b.
	The arrows in Fig.~\ref{fig:lat_dv}b show the magnetization of the neighboring islands in the configuration $c=63$.
	\subsection{\label{sec:level3_B}MODEL~1.~MEAN-BARRIER APPROXIMATION}
	For the first model, we introduce the NN islands in our modeling.
	Motivated by the symmetry properties of the dipolar interaction and Zeeman energies, the mean-barrier\cite{Farhan2013,farhan_prl,farhanderlet2017,arava2019,Thonig2014} of the switching is described by 
	\begin{equation}
	\centering
	\label{eq:de_1}
	{\Delta E^\mathrm{mean}} = \Delta E^{\mathrm{isol}} + \frac{1}{2}{\left(E^{\mathrm{fin}} - E^{\mathrm{ini}}\right)} ,
	\end{equation}
	where $\Delta E^{\mathrm{isol}}$ is the energy barrier to switch the magnetization of one isolated nanostructure, $E^{\mathrm{fin}}$ the energy of the final state corresponding to the switched double vertex and $E^{\mathrm{ini}}$ the initial energy of the double vertex.
	A derivation of Eq.~(\ref{eq:de_1}) is given in Appendix~\ref{sec:derivation_mean_barrier}.
	The magnetizations of the neighbors as well as of the central island are kept uniform during the switching event.
	Since only the initial and final state are involved in Eq.~(\ref{eq:de_1}), we do not need to apply the SIMS in this model. 
	In order to keep each island uniformly magnetized and the reversal mechanism a coherent rotation, we consider the energy barrier for one single nanostructure $\Delta E^{\mathrm{isol}}$, which is obtained from a coherent rotation of the magnetization.
	Thus, this yields $\Delta E^{\mathrm{isol}}=\SI{1.37}{eV}$.
	\subsection{\label{sec:level3_C}MODEL~2.~UNIFORM ENVIRONMENT}
	As a step further, we apply the SISM in this model in such a fashion that $\Omega_{m}$ covers only this island.
	We continue to keep the NN islands uniformly magnetized.
	Under these assumptions, the total energy of the system also includes the dipolar interactions between the central island and the NN islands, as in the previous model.
	The interaction energies among the NN islands are constant during the switching process due to their uniform nature.
	Therefore, the difference in their saddle point energy and initial energy vanishes.
	Note that the reversal mechanism is not a perfect coherent rotation anymore.
	\subsection{\label{sec:level3_E}MODEL~3.~FULL MICROMAGNETIC MODEL}
	For the last approach, we include all the seven islands in the magnetic region $\Omega_m$, where the string method is applied.
	Thus both the NN islands and the central island are dynamically relaxed.
	The interactions among the NN islands change at each state of the minimum energy path, so they cannot be neglected in the barrier calculation.
	
	We include indirectly the effect of temperature in our system via reduced $M_s$ and $A^\mathrm{ex}$~\cite{heider88,martinez2003,evans2016}, allowing for fluctuations to arise from the decreased coupling energy between the spins, which could even form magnetic domains.
	In Model~3, fluctuations are accounted for not only in the central island but also in the NN islands.
	Note that Model~2 also considers for the fluctuations, however only in the magnetization of the central island, since the NN islands are kept uniform.  
	\section{\label{sec:level4}RESULTS}
\begin{figure}[]
		\includegraphics[width=\columnwidth]{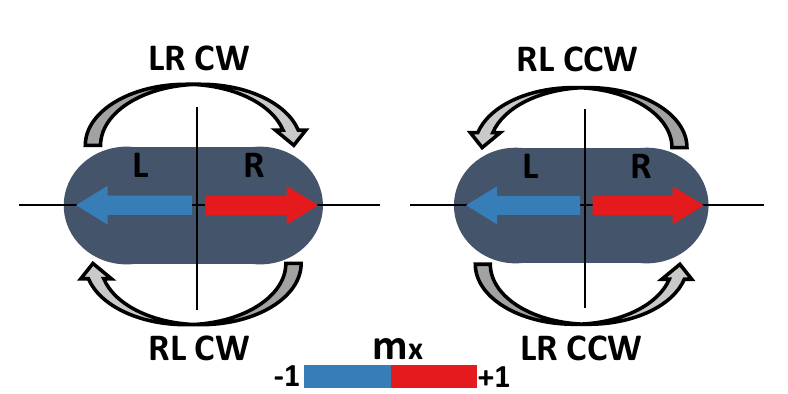}
		\caption{\textbf{Rotation directions for the switching event on the central island.} We refer to a switching event with $\boldsymbol{m}^{\mathrm{ini}}~=~(-1,0,0)$, $\boldsymbol{m}^\mathrm{saddle}_\mathrm{CCW}~=~(0,-1,0)$ and $\boldsymbol{m}^{\mathrm{fin}}~=~(1,0,0)$ as a left to right counterclockwise rotation~(LR~CCW).
			If the saddle point configuration is changed to $\boldsymbol{m}^\mathrm{saddle}_\mathrm{CW}~=~(0,1,0)$, this becomes a left to right clockwise~(LR~CW) rotation.
			Likewise, one can define a right to left counterclockwise rotation~(RL~CCW) and a right to left clockwise rotation~(RL~CW).}
		\label{fig:binary}
	\end{figure}

	\begin{table*}[t]
		\caption{\textbf{Energy barriers and estimated switching times obtained by applying the Models~1-3}. The directions are as visualized in Fig.~\ref{fig:binary}. The estimated switching times were calculated with Eq.~(\ref{eq:tau}) using $\tau_0=\SI{e-10}{s}$ and $T=\SI{300}{K}$.}
		\begin{tabular}{ccccccc}
			\toprule
			\toprule
			& Isolated & \multicolumn{5}{c}{Configuration $c=63$}\\ 
			\midrule
			Direction & \multicolumn{2}{c}{CCW/CW} & \multicolumn{2}{c}{LR~CCW} & \multicolumn{2}{c}{LR~CW}\\
			\midrule
			MODEL No.& - & 1 & 2 & 3 & 2 & 3 \\
			\midrule
			$\Delta E~\mathrm{(eV)}$ & 1.37 & 1.16 & 1.51 & 1.37 & 0.83 & 0.77\\
			$\tau~(\mathrm{s})$ & $\num{9.8e12}$ & $\num{2.4e9}$ & $\num{2.9e15}$ & $\num{1.2e13}$ & $\num{7.2e3}$ & $\num{8.4e2}$\\  
			\bottomrule
			\bottomrule
		\end{tabular}
		\label{tab:values}
	\end{table*}
	One of the main consequences introduced by Model~2 and Model~3 and that is not predicted by Model~1, is the dependence of the energy barrier on the direction of the rotation of the central island's magnetization. 
	For the sake of simplicity, we give to each rotation direction an acronym, as illustrated in Fig.~\ref{fig:binary}. 
	To analyze the energy barriers in sASI systems, we compare the models from Section~\ref{sec:level3}.
	First we consider a single configuration $c=63$, as illustrated in Fig.~\ref{fig:lat_dv}b.
	In this case, a left to right~(LR) switching of the magnetization of the central island changes the state of the double vertex from a Type~III-Type~III double vertex to an energetically favourable Type~II-Type~II.
	Thereby, the switching of the central island's magnetization can occur via two parallel channels, CW and CCW, which might have a different switching rate depending on the NN configuration.
	The resulting rate of switching from the Arrhenius law\cite{brown63,Coffey2012} is then defined as
	\begin{equation}
	\label{eq:rates}
	f=\dfrac{1}{\tau}=\dfrac{1}{\tau_0}\cdot \left( e^{-\dfrac{\Delta E_\mathrm{CW}}{k_BT}}+e^{-\dfrac{\Delta E_\mathrm{CCW}}{k_BT}} \right),
	\end{equation}
	with $\tau$ being the estimated lifetime of the state, $\tau_0$ the attempt period, $k_{\mathrm{B}}~=~\SI{8.62e-5}{eV/K}$ the Boltzmann constant and $T$ is the temperature.
	When a mean barrier approximation is used, the switching rate is given as
	\begin{equation}
	\label{eq:rates2}
	f^\mathrm{mean}=\dfrac{2}{\tau_0}\cdot  e^{-\dfrac{\Delta E^\mathrm{mean}}{k_BT}},
	\end{equation}
	where $\Delta E^\mathrm{mean}$ is the mean barrier given by Eq.~(\ref{eq:de_1}) which is the average barrier between the CW and CCW barriers.
	However, in the cases where the rates via the CW and CCW channels are very different, only the fastest channel will dominate. Thereby, the mean-barrier approximation leads to significantly deviating rates. A detailed discussion for these deviations is given in Appendix~\ref{sec:derivation_mean_barrier}.
	Under these assumptions, Eq.~(\ref{eq:rates}) can be simplified including only the exponential term corresponding to the faster channel. Hence, the estimated lifetime for a single channel is defined by
	\begin{equation}
	\label{eq:tau}
	\tau={\tau_0}\cdot  e^{\dfrac{\Delta E}{k_BT}}.
	\end{equation}

\begin{figure}[]
		\includegraphics[width=\columnwidth]{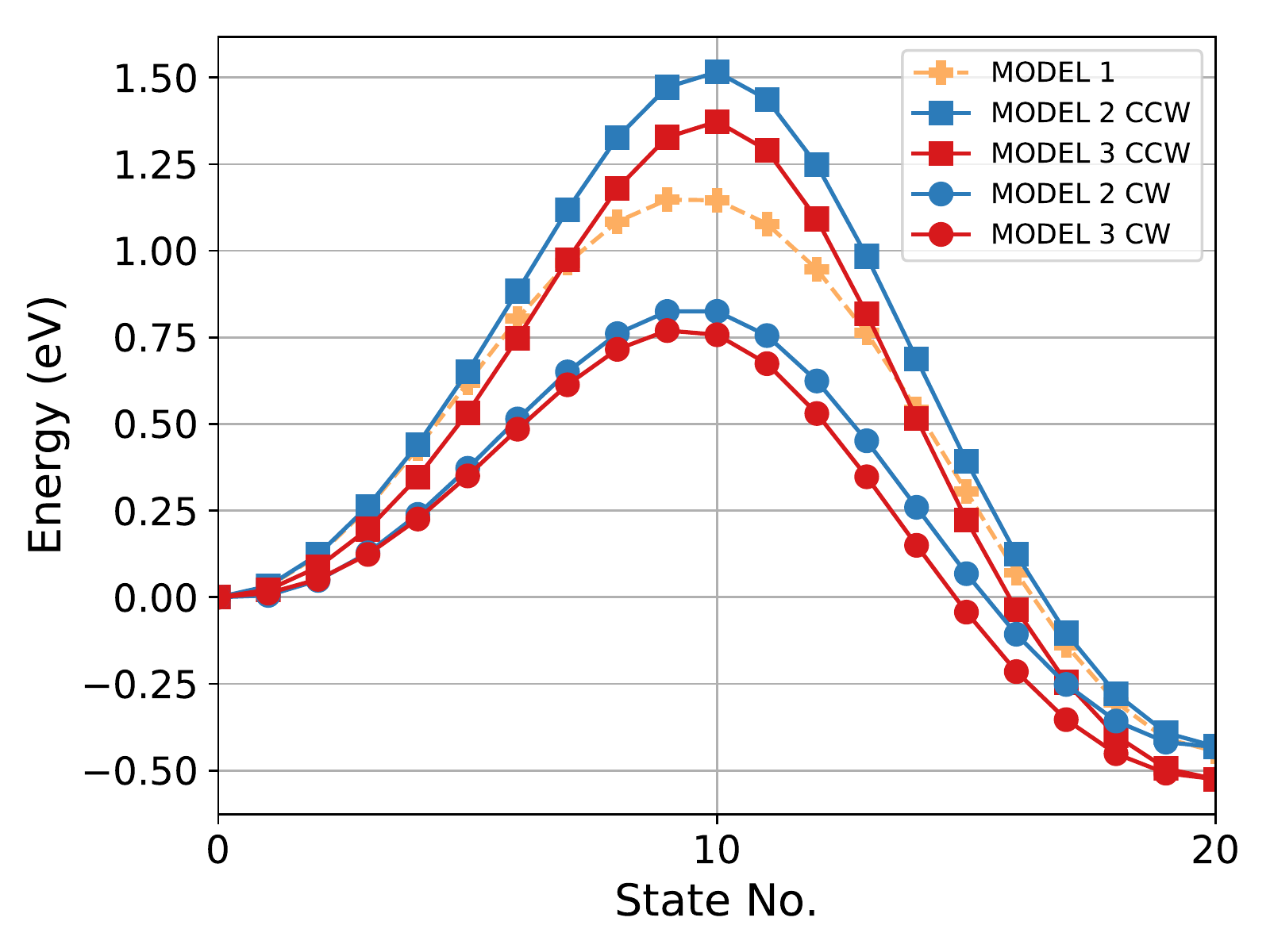}
		\caption{\textbf{Minimum energy paths for the switching of the magnetization of the central island.} The MEPs for $c=63$ considering both CCW~(squares) and CW~(circles) rotations, where the initial energies are shifted to zero. Here, the initial double vertex contains two Type~III vertices and one obtains two Type~II vertices after switching the magnetization of the central island from $\boldsymbol{m}^{\mathrm{ini}}=(-1,0,0)$ to $\boldsymbol{m}^{\mathrm{fin}}=(1,0,0)$, represented by State No. = 0 and State No.~=~20 respectively.}    
		\label{fig:direc}
	\end{figure}

	Fig.~\ref{fig:direc} illustrates the MEPs for both CW and CCW rotations, where the initial energies have been shifted to zero (see Tab.~\ref{tab:values}).
	The results indicate that the energy barriers for a CW rotation are lower than for a CCW rotation.
	As the name suggests, Model~1 is the average barrier for the CW and CCW rotations, and thus, the energy path from Fig.~\ref{fig:direc} is obtained as the average energy path for these directions.
	According to Eq.~\ref{eq:de_1}, the energy barrier for Model~1~(orange) depends only on energies of the initial and final states and does not distinguish between the rotation directions.
	Models~2 and 3 (blue and red), however, include the spatial distribution and symmetry of the neighboring islands in the minimization process during the string method, thus breaking this degeneracy in some cases.
	Namely, when considering the demagnetization field interactions from the neighboring islands, the CW rotation is energetically favorable over the CCW rotation since its intermediate magnetization configurations point in positive y-direction which aligns with the non-zero y-component of the strayfield in the  configuration $c=63$.
	Table \ref{tab:values} shows the values for the energy barriers of the configuration $c=63$ and of the isolated island.

	In the following, we focus on the CW rotations for the configuration $c=63$.
	All models show a significant reduction of the energy barrier compared to the isolated nanostructure mainly due to the introduction of the dipolar interaction with the NN islands.
	In contrast to Model~1,  Model~2 introduces non-uniformities in the magnetization of the central island.
	The reversal mechanism obtained with the string method, which is not a perfect coherent rotation anymore, is associated to a further reduction of the energy barrier.
	Compared to the widely used mean-barrier approximation~(Model~1), the Model~2 reduces the lowest energy barrier~(LR~CW) about 28\%.
	The highest energy barrier reduction is observed for Model~3, which is the full micromagnetic model.
	In addition to Model~2, the neighbors are included in the magnetic region of the SISM, where the energy is minimized.
	Thus their magnetizations are dynamically relaxed and may change for each intermediate state in the transition path.
	In this case, the dipolar interactions among the neighboring islands are not constant and the difference of their saddle point and initial states energies does not vanish.
	The lowest energy barrier for this model is reduced by 33\% compared to Model~1.
	Furthermore, only for Model~3 the magnetic system is in a true energetic minimum, since all islands are added in the magnetic region of the string method and their energies have been minimized.
	Animations of the switching process for the configuration $c$ using Model~2 and Model~3 can be found in the supplementary materials.
	
	Although the differences between the Models~2~and~3 may seem rather small, they have a major impact on the average lifetime given by Eq.~\eqref{eq:tau}.
	Since the fastest channel dominates the switching, we compare the estimated switching times for LR~CW rotation directions for the configuration $c=63$.
	Here we use the attempt period $\tau_0=\SI{e-10}{s}$ and $T=\SI{300}{K}$.
	The value of the attempt period is only exemplary to show the role of the energy barrier reduction regarding the switching times.
	The values for $\tau$ are given in Tab.~\ref{tab:values}.
	One can see that the estimated lifetime~(LR~CW) calculated using Model~2 is approximately up to six orders of magnitudes lower compared to Model~1, and one order of magnitude lower with respect to Model~3.
	The full micromagnetic model yields a reduction of the average lifetime by about seven orders of magnitude with respect to the average barrier Model 1, often used in ASIs literature.
	
	To show that Model~3 predicts (in nearly all cases) the lowest energy barriers and switching rates, we calculated the switching barriers and associated rates for all possible configurations $c$ in a double vertex given by Eq.~(\ref{eq:c_i}).
	\begin{figure}[]                               
		\begin{subfigure}{\columnwidth}
			\includegraphics[width=0.8\columnwidth]{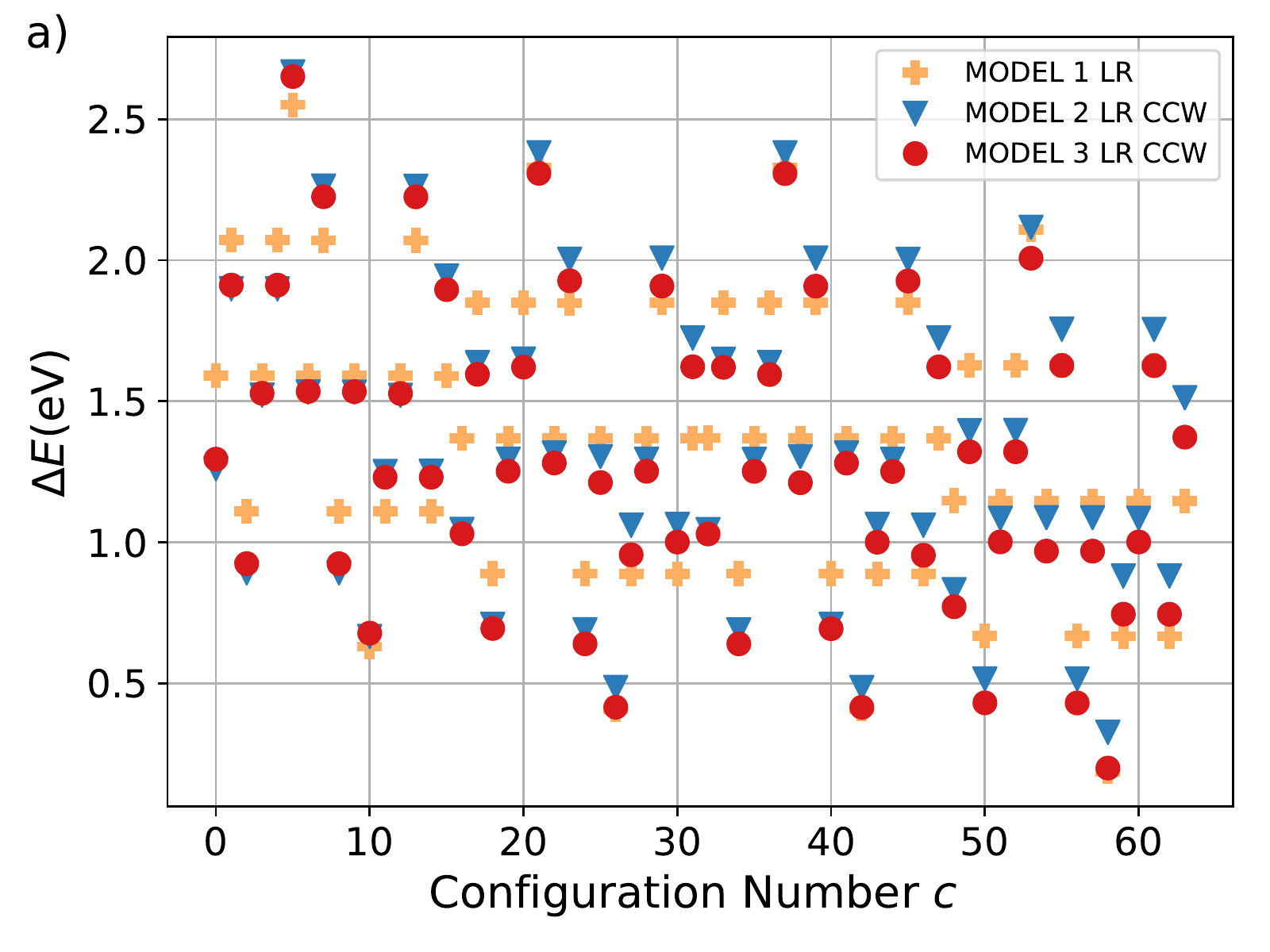}
		\end{subfigure}
		\begin{subfigure}{\columnwidth}
			\centering	
			\includegraphics[width=0.8\columnwidth]{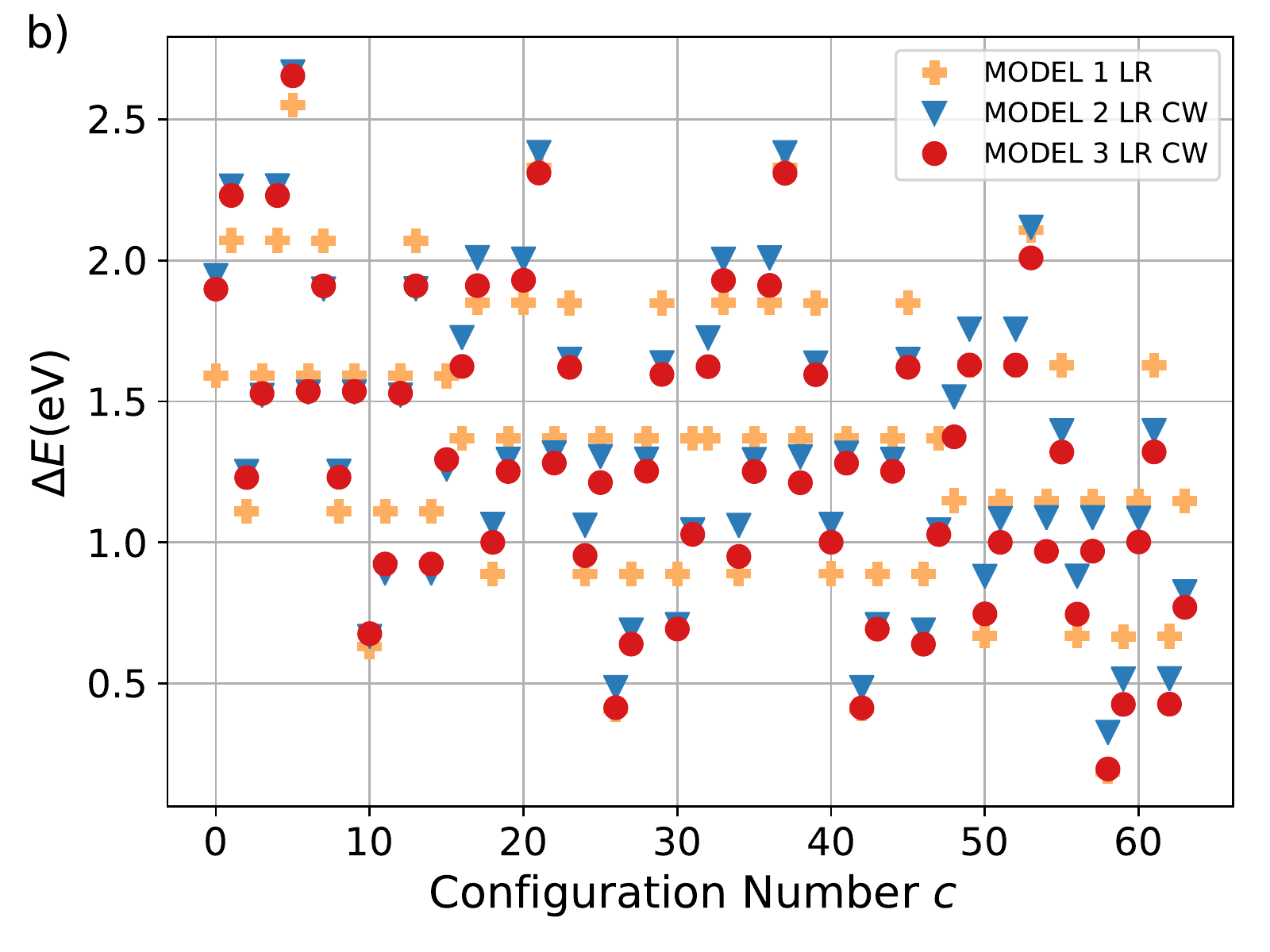}
		\end{subfigure}
		\caption{\textbf{Direct comparison between the Models.} Energy barriers for the switching of the magnetization of the central island using all possible configurations of the NN islands in a double vertex illustrated in Fig.~\ref{fig:lat_dv}b via LR~CCW~(a) and LR~CW~(b) rotation directions.}
		\label{fig:barriers}
	\end{figure}
	Fig.~\ref{fig:barriers} shows a direct comparison between the models for both LR~CCW and LR~CW rotations.
	To obtain the barriers with respect to the final state (RL) one has to rotate the  islands $180^\circ$ around both the vertical and horizontal axis, since the rotation direction also changes.
	In principle there exists an equivalent CCW barrier for all CW barriers, but with a different configuration number.
	If we consider $c=63$ as an example, $c=0$ LR~CCW has the same barrier as $c=63$ RL~CCW.
	
	Fig.~\ref{fig:barriers} shows that for all configurations at least one of the possible energy barriers is the lowest for Model~3.
	Note that Model~1 as well as Model~2 can result, in some cases, in lower energy barriers compared to Model~3.
	This occurs due to the fact that the initial and final states are not properly  minimized, and thus the system is not in equilibrium, i.e. an equal reduction of the saddle point energy for both Model~2 and Model~3, can lead to a lower energy barrier for Model~2.
	Since Model~3 describes the evolution of the system through true energy minima, it yields the most realistic barrier.

	\begin{figure}[]
		\includegraphics[width=\columnwidth]{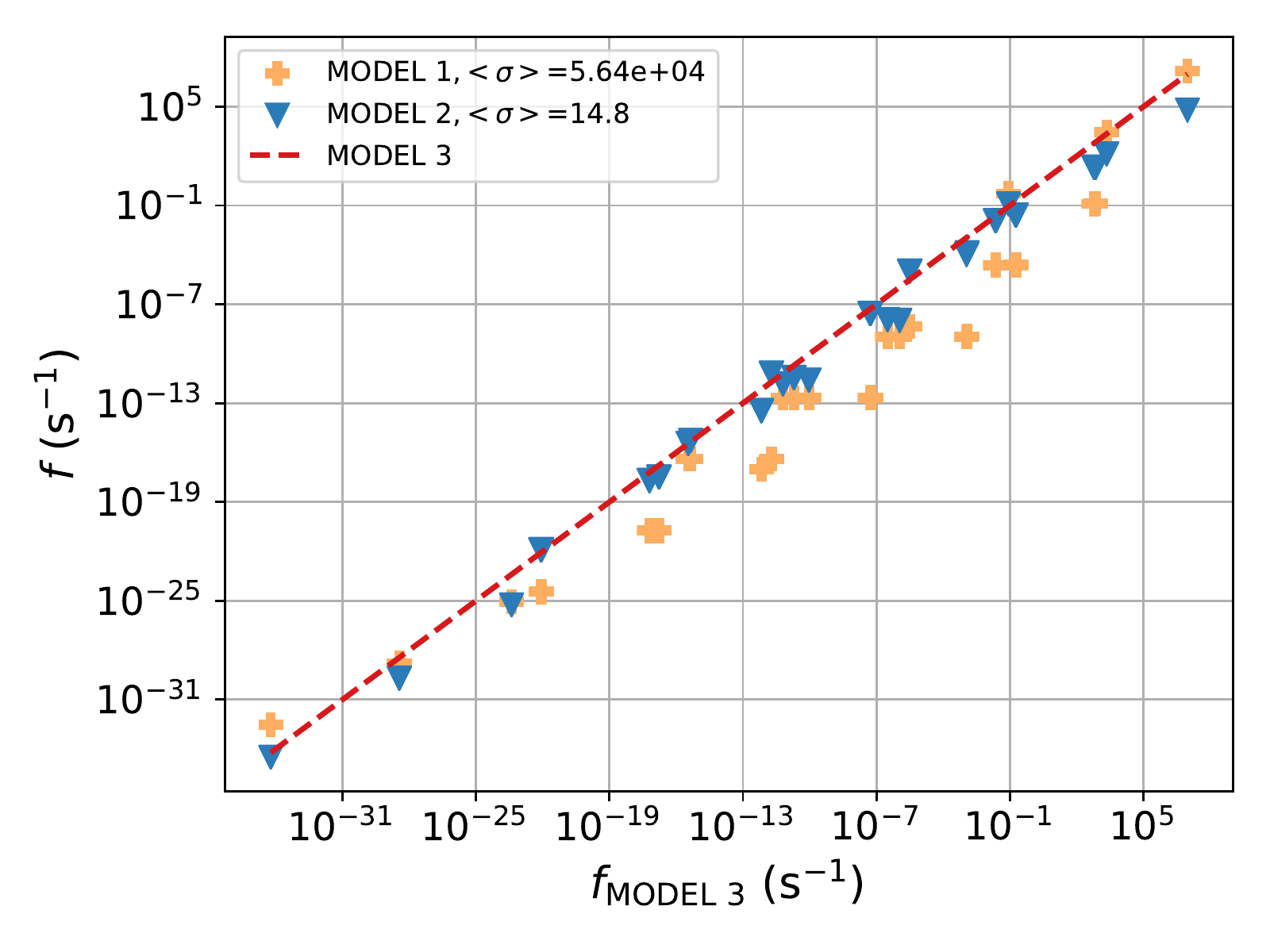}
		\caption{\textbf{Switching rates for the Models~1-3.} Log-log plot of the switching rates given by Eq.~(\ref{eq:rates}) for each model. The x-axis shows the switching rates for the possible 64 configurations using Model~3, whereas the y-axis shows the switching rates for the Models~1~(orange, crosses) and 2~(blue, triangles). The mean error $<\sigma>$, given by Eq.~\ref{eq:sigma}, is shown in the legend. The values of Model~3 are plotted as a reference line.}
		\label{fig:rates}
	\end{figure}
	Fig.~\ref{fig:rates} illustrates the comparison of the switching rates calculated with Eq.~(\ref{eq:rates}) and (\ref{eq:rates2}) using the energy barriers obtained with Models~1~and~2 with respect to Model~3. 
	The mean error
	\begin{equation}
	\centering
	\label{eq:sigma}
	<\sigma>=\frac{1}{64}\sum_c\frac{f_{3,c}}{f_{i,c}},
	\end{equation}
	where the value $\sum_c{}{\frac{f_{3,c}}{f_{i,c}}}$ sums over the relative deviation factors for all configurations $c$, gives an estimate of the average deviation from the full micromagnetic model.
	While $<\sigma>=1$ means a perfect agreement on average, the higher the deviation from this value, the more inaccurate are the calculated energy barriers.

	Even though Model~1 might be a valid approximation of the mean-barrier based on the derivation given in Appendix~\ref{sec:derivation_mean_barrier}, the concept of an average barrier does not necessarily imply a physically justifiable input for the kinetic Monte Carlo simulations. Considering once more the configuration $c=63$, we calculate the exact average barrier using the values obtained from Model~3, hence, the average barrier is $\Delta E^\mathrm{avrg}={1/2(0.77 + 1.37)}~{\mathrm{eV}}=\SI{1.07}{eV}$. For the switching rates of the configuration $c=63$ calculated with Eq.~\eqref{eq:rates} and \eqref{eq:rates2} we obtain $f=\SI{1.2e-3}{s^{-1}}$ and $f^\mathrm{avrg}=\SI{2.2e-8}{s^{-1}}$. Compared to $f^\mathrm{mean}$ obtained with the mean-barrier from Model~1, which is $f^\mathrm{mean}=\SI{6.6e-10}{s^{-1}}$, we see that the proper micromagnetic modeling $f^\mathrm{mean} \rightarrow f^\mathrm{avrg}$ leads to an increase of the switching rate up to two orders of magnitudes, whereas differentiation between parallel switching channels $f^\mathrm{avrg} \rightarrow  f$ improves the switching rate by additional five orders of magnitude.
	
	In summary, our results indicate that the values obtained with Model~1 deviate significantly from the values of Model~3.
	The mean barrier approximation will falsely overestimate the energy barriers in the cases, where an interaction field with a non-vanishing y-component acts on the central island, and thus, the CW and CCW barriers are different. As a consequence, the switching rates are underestimated.
	Although Model~2 recovers the most important shortcomings of Model~1 by using micromagnetics and differentiating between the clockwise and counterclockwise channels, it is Model~3 that gives the lowest and more realistic energy barriers.
	Thereby, Model~3 provides the best approximation of the energy barriers and transition rates that should be utilized to model dynamical processes in sASI lattices, since the probability of switching the magnetization of an island is directly proportional to these rates.
	\section{\label{sec:level5}CONCLUSIONS}
	In this paper we use the simplified and improved string method to calculate all energy barriers to switch the magnetization of the central island of a double vertex in square artificial spin ice.
	We investigate the influence of the nearest neighbor islands on the energy barrier in a square artificial spin ice lattice by calculating the energy barrier with three different approaches.
	In the first model we consider the widely used mean-barrier approximation.
	Besides the non-uniformities in the magnetization of the island of interest considered in Model~2, the last model, Model~3, is a full micromagnetic model where each magnetization is dynamically relaxed depending on each other.
	As a first relevant result, Model~2 yields different minimum energy paths for counterclockwise and clockwise rotation directions for particular configurations, where the most probable switching occurs via the channel with the lowest energy barrier.
	This distinction between clockwise and counterclockwise reversal is completely neglected by the average model, Model~1, often utilized in ASIs literature.
	
	To conclude our results, the energy barrier for switching the magnetization of an island in an artificial spin ice lattice can be reduced significantly applying a full micromagnetic model on a double vertex~(Model~3) compared to the energy barrier obtained with the mean-barrier approximation.
	The interactions originating from the demagnetization fields from the neighboring islands and the consequent energy difference between clockwise and counterclockwise reversal paths are the first key reason for this reduction.
	The inhomogeneities in the magnetizations of both the central islands and of the dynamically relaxed neighbors, that arise during the reversal, introduce an additional contribution to the reduction of the energy barriers.
	Both effects are neglected in the mean-barrier approximation.
	In most drastic cases the mean-barrier-approximation can result in energy barriers $35\%$ higher as Model~3.
	While the \textit{ad hoc} reduction of the barriers often applied in mean-barrier model can fix particular barriers, it can over- or underestimate the barriers leading to different dynamics in artificial spin ice systems. The main reason being the fact, that an average barrier has no physical significance regarding the switching rates utilised in the kinetic Monte Carlo simulations.
	
	The presented full micromagnetic model is general and can also be applied for other artificial spin ice lattice types and magnetic materials, if both the number and spatial arrangement of the nearest neighbors are adapted.
	\appendix
	\section{Mean-barrier approximation}
	\label{sec:derivation_mean_barrier}

	Especially for kinetic Monte Carlo simulations of artificial spin ice, the mean switching barrier is often expressed \textit{solely} depending on the energies of the initial and final energy state. The initial and final state energies can be easily calculated using the point-dipole model or regular micromagnetic simulation codes, where each island is uniformly magnetized, with a single-island switching barrier $E^{\mathrm{isol}}$ added as an independent parameter.
	The reason that the energies of the intermediate states do not need to be considered in this simple approximation is based on the symmetry properties of the dipolar interaction, respectively Zeeman energy, which are antisymmetric under rotations of $\pi$, i.e.\ $E(\phi + \pi) = -E(\phi)$. Under the assumptions of a Stoner Wohlfarth particle as central island and constantly magnetized neighboring islands, the total energy of the central island is given by an anisotropy and Zeeman term in the form
	\begin{equation}
	E(\boldsymbol{m},\boldsymbol{H}) = E^\text{ani}(\boldsymbol{m}) + E^\text{zee}(\boldsymbol{m},\boldsymbol{H}),
	\end{equation}
	with
	\begin{align}
		E^\text{ani}(\boldsymbol{m}) &= -K_V(\boldsymbol{m} \cdot (1,0,0))^2,\\ 
		E^\text{zee}(\boldsymbol{m},\boldsymbol{H}) &= -\mu_0 M_s\boldsymbol{m}\cdot \boldsymbol{H},
	\end{align}
	where $K_V$ denotes the product of the effective anisotropy constant $k_\mathrm{eff}$ and the volume of the isolated nanostructure.
	To obey the aforementioned symmetry argument, $\boldsymbol{H}$ must be assumed as homogeneous. Furthermore, we assume that this field is sufficiently weak, so that both the initial and final states can be approximated by $\boldsymbol{m}^\text{ini} = (-1,0,0)$ and $\boldsymbol{m}^\text{fin} = (1,0,0)$. In general, the value of the activation barrier is given by the difference between the saddle point energy $E^\mathrm{saddle}$ as given by Eq.~(\ref{eq:de_tst}), i.e.\ central moment pointing up, $E^\mathrm{saddle}_{\mathrm{CW}}$ with $\boldsymbol{m}^\mathrm{saddle}_\mathrm{CW}~=~(0, 1, 0)$, or down, $E^\mathrm{saddle}_{\mathrm{CCW}}$ with $\boldsymbol{m}^\mathrm{saddle}_\mathrm{CCW}~=~(0, -1, 0)$ in a configuration $c$, and the initial energy of the configuration $E^\mathrm{ini}$. Note, that the magnetization state with the maximum energy involved in the energy barrier calculation can significantly deviate from $\boldsymbol{m}^\mathrm{saddle}_\mathrm{CW/CCW}$ for high interaction fields. We assume that this deviation is also negligible.
	In this context we obtain:
	\begin{align}
		E^\text{ini}(\boldsymbol{H}) &= E(\boldsymbol{m}^\text{ini},\boldsymbol{H}) = - K_V + \mu_0 M_s H_x\\
		E^\text{fin}(\boldsymbol{H}) &=E(\boldsymbol{m}^\text{fin},\boldsymbol{H}) = - K_V - \mu_0 M_s H_x\\
		E^\text{saddle}_\text{CW}(\boldsymbol{H}) &=E(\boldsymbol{m}^\mathrm{saddle}_\mathrm{CW},\boldsymbol{H}) =  - \mu_0 M_s H_y\\
		E^\text{saddle}_\text{CCW}(\boldsymbol{H}) &=E(\boldsymbol{m}^\mathrm{saddle}_\mathrm{CCW},\boldsymbol{H}) =  + \mu_0 M_s H_y
	\end{align}
	We get for the energy barrier of an isolated nanostructure
	\begin{align*}
		\Delta E^{\mathrm{isol}} = E^\mathrm{saddle}_{\mathrm{CW}} (\boldsymbol{H}=(0,0,0)) - E^\text{ini}( \boldsymbol{H}=(0,0,0)) = K_V.
	\end{align*}
	
	With this result, we are finally able to express the mean-energy barrier of the CW and CCW channels by $\Delta E^\mathrm{isol}, E^\text{fin}$ and $E^\text{fin}$ as in Eq.~\eqref{eq:de_1}:
	\begin{align*}
		\Delta E^\mathrm{mean} &= 1/2 \big(\Delta E^\text{saddle}_\text{CW} + \Delta E^\text{saddle}_\text{CCW} \big)\\
		&= 1/2  \big( E^\text{saddle}_\text{CW}(\boldsymbol{H}) - E^\text{ini} (\boldsymbol{H}) + E^\text{saddle}_\text{CCW} - E^\text{ini} (\boldsymbol{H}) \big)\\
		&= K_V - \mu_0 M_s H_x \\ 
		&=\Delta E^\mathrm{isol} + 1/2 \big( E^\text{fin}(\boldsymbol{H}) - E^\text{ini}(\boldsymbol{H}) \big).
	\end{align*}
Based on the aforementioned assumptions, Eq.~(\ref{eq:de_1}) is an approximation of an average barrier between the two energy barriers for the corresponding parallel switching channels. Even when the initial and final states are involved in this equation, it might neglect the interaction fields originating from the neighboring islands. In a case, where $E^\mathrm{ini}=E^\mathrm{fin}$, the mean-barrier coincides with the energy barrier of one isolated nanostructure, e.g. configuration c=31. But there is still an effective field acting on the central island. According to the Stoner Wohlfarth Model, the energy barrier of switching the magnetization of a single domained particle under the influence of an external field $\boldsymbol{H}=(0,H_y,0)$ is $\Delta E = \Delta E^{\mathrm{isol}}\left(1-\dfrac{H_y}{H_k}\right)^2$, where $H_k$ is the strength of the anistropy field and $H_y$ the strength of the field acting on the single-domained particle. With Eq.~(\ref{eq:de_1}) one would obtain that $\Delta E^\mathrm{mean}=\Delta E^{\mathrm{isol}}$. This deviation points out once more, that the mean-barrier approximation method needs further corrections, to be used as input for kinetic Monte Carlo simulations.

	\begin{acknowledgments}
		The computational results presented have been achieved [in part] using the Vienna Scientific Cluster (VSC).
		S.K., C.A., C.V., F.S., F.B., P.H. and D.S. gratefully acknowledge the financial support by the Austrian
		Federal Ministry for Digital and Economic Affairs and the
		National Foundation for Research, Technology and Development.
		N.L., M. M., and P.V. acknowledge support from the Spanish Ministry of Economy, Industry and Competitiveness under the Maria de Maeztu Units of Excellence Programme - MDM-2016-0618 and the project RTI2018-094881-B-I00 (MINECO/FEDER).
		N.L. has received funding from the European Union’s Horizon 2020 research and innovation programme under the Marie Skłodowska Curie grant agreement No.~844304 (LICONAMCO).
		K.H. acknowledges funding by the Swiss National Science Foundation (Project No.~200020{\_}172774).
		
	\end{acknowledgments}
	\bibliography{manuscript}
\end{document}